\documentclass[aip,reprint]{revtex4-2}

\usepackage{graphicx}
\usepackage{dcolumn}
\usepackage{bm}
\usepackage{xr}
\usepackage{gensymb}
\usepackage{amsmath}
\usepackage{siunitx}

\usepackage{comment}

\usepackage[intoc]{nomencl}
\usepackage[hidelinks]{hyperref}

\begin{document}

\title{Improvement of Data Analytics Techniques in Reflection High Energy Electron Diffraction to Enable Machine Learning}

\author{Patrick T. Gemperline}
\affiliation{Department of Physics, Auburn University, Auburn, AL, USA}

\author{Rajendra Paudel}
\affiliation{Department of Physics, Auburn University, Auburn, AL, USA}

\author{Rama K. Vasudevan}
\affiliation{Center for Nanophase Materials Science, Oak Ridge National Laboratory, Oak Ridge, TN, USA}

\author{Ryan B. Comes}
\affiliation{Department of Physics, Auburn University, Auburn, AL, USA}
\affiliation{Department of Materials Science and Engineering, University of Delaware, Newark, DE, USA}
\email{comes@udel.edu}

\date{\today}

\begin{abstract} 
Perovskite oxides such as LaFeO$_3$ are a well-studied family of materials that possess a wide range of useful and novel properties. Successfully synthesizing perovskite oxide samples usually requires a significant number of growth attempts and a detailed film characterization on each sample to find the optimal growth window of a material. The most common real-time \textit{in situ} diagnostic technique available during molecular beam epitaxy (MBE) synthesis is reflection high-energy electron diffraction (RHEED). Conventional use of RHEED allows a highly experienced operator to determine growth rate by monitoring intensity osciallations and make some qualitative observations during growth, such as recognizing the sample has become amorphous or recognizing that large islands have formed on the surface. However, due to a lack of theoretical understanding of the diffraction patterns, finer, more precise levels of observations are challenging. To address these limitations, we implement new data analytics techniques in the growth of three LaFeO$_3$ samples on Nb-doped SrTiO$_3$ by MBE. These techniques improve our ability to perform unsupervised machine learning using principal component analysis (PCA) and k-means clustering by using drift correction to overcome sample or stage motion during growth and intensity transformations that highlight more subtle features in the images such as Kikuchi bands. With this approach, we enable the first demonstration of PCA and k-means across multiple samples, allowing for quantitative comparison of RHEED videos for two LaFeO$_3$ film samples. These capabilities set the stage for real-time processing of RHEED data during growth to enable machine learning-accelerated film synthesis.\end{abstract}
\maketitle


\section {Introduction}

    Molecular beam epitaxy (MBE) is a thin film synthesis technique that is able to achieve highly crystalline epitaxial growth of single crystal samples. In order to determine the underlying structure and properties of thin films synthesized using MBE, several different characterization techniques are used. In most systems, due to the geometry of the chambers and the temperature and pressure conditions, RHEED is the only \emph{in situ} characterization technique possible. During film growth, \emph{in situ} RHEED patterns provide information on the evolution of the sample in real time rather than after growth~\cite{RHEED_epitaxy}. This provides opportunities for the operator to adjust growth conditions during growth to improve sample quality. In addition to monitoring the RHEED patterns of the film, there is information that can be gained by monitoring the intensity of the RHEED spots themselves. It is common to use a program to sum the intensity within a certain region or a single spot and plot this over time~\cite{RHEED_book}. When using this method, oscillations can be observed in the intensity of the specular spot, which are indicative of the sample undergoing layer-by-layer growth~\cite{RHEED_book}. Since this cycle corresponds to the deposition of a single unit cell, it is possible to precisely tune a growth to obtain a desired number of unit cells or even end in a particular surface termination. 

    Although RHEED is highly useful during film synthesis, there are several downsides and limitations. Aside from the determination of film growth rates, the results extracted from the RHEED patterns are limited to primarily qualitative observations rather than quantitative ones. Thus, the first and foremost limitation is that it requires a well-trained expert to comprehend and interpret the patterns. Without sufficient experience, recognizing the patterns and the subtle changes during growth renders the technique much less useful. This experience takes many growths over months or years to acquire. Significant experience is needed to analyze RHEED images because there are few advanced analytical tools that can be used to analyze the images. While theory is able to successfully explain the location of the Laue rings and RHEED spots~\cite{RHEED_Spots}, this is possible under ideal circumstances and ignores the other effects present in real samples. 

    Machine learning is already a highly developed area of research and has been applied to a wide variety of image processing tasks. Handwriting recognition~\cite{Hand_writting} and generation~\cite{cGAN} are some of the most common. In materials research, machine learning techniques have been applied to a wide array of material characterization and analysis techniques, including STEM~\cite{STEM1}, SPM~\cite{SPM}, transport measurements~\cite{Transport}, surface morphology~\cite{Rama_surface}, and crystal structure determination~\cite{Crystal_determination}. Efforts in RHEED have expanded dramatically in recent years, with initial efforts focusing on unsupervised learning approaches, including principal component analysis (PCA), k-means clustering, and negative matrix factorization (NMF) \cite{RAMA_RHEED, Sydney, Suyolcu2021, gliebe2021distinct}. Several studies have also been conducted that focus on utilizing machine learning techniques such as neural networks and random forests to make predictive models that utilize various forms of data, including XANES spectra~\cite{XANES_ML}, materials structure databases~\cite{Crystal_Predictions}, and MBE operation data~\cite{MBE_ML}, to extrapolate additional information on the  structural and chemical qualities of materials. Neural networks have also been applied to RHEED data as a means to predict film stoichiometry \cite{price2024predicting}, surface deoxidation \cite{khaireh2023monitoring}, and azimuthal alignment of the substrate \cite{kwoen2020classification}.
    
    In this work, principal component analysis and $k$-means clustering are used on recordings of RHEED patterns to provide quantitative feedback that informs on the evolution of the film surface during growth. Of note, past work has explored the application of modern ML methods to RHEED image sequences, but has not seriously engaged with the problem of how to normalize and pre-process RHEED datasets arising from growths of different types of thin film samples. This is particularly important if RHEED is to be used in the context of future autonomous synthesis platforms, where reliable in situ monitoring that can be normalized and standardized for future use is of prime importance. Here, we explore several methods for normalizing intensity, structure, and geometry of RHEED image sequences to ready them for ML analysis. We show that it is possible to make comparisons across samples in a systematic fashion using these approaches by obtaining a common set of principal components and centroid images that describe multiple samples. This capability allows for comparison of material similarity between samples in a systematic fashion.

\section{Methods}
    \subsection{Film Growth}
        LaFeO$_3$ (LFO) films were grown using plasma-assisted oxide molecular beam epitaxy as described previously~\cite{Burton2022}. Metallic La and Fe sources were heated to obtain atomic fluxes of each element and were calibrated prior to growth using a quartz crystal microbalance (QCM). (001) Nb-doped SrTiO$_3$ (STO) substrates (MTI Corporation) were used for all growths. Prior to growth, the substrates were heated to 700~$\degree$C in \num{3e-6}~Torr oxygen background pressure. A Mantis RF oxygen plasma source operating at 300~W forward power was used to increase the concentration of atomic oxygen. Growth rates for LFO were typically 45-90 seconds per formula unit, producing an overall growth rate of approximately 3-6~\r{A}/min. Film growth was monitored via RHEED using a Staib Instruments electron gun operated at 15~kV accelerating voltage. Videos of the patterns were captured via Flashback screen recording software, as described previously~\cite{Sydney}. Videos were saved for post-growth analysis by several different data analytics techniques.
    
    \subsection {Data Preprocessing}

        Several different methods of preprocessing were used to prepare RHEED videos for use in the PCA and $k$-means algorithms. PCA is a computationally intensive algorithm that scales with the number of samples, $n$, and the dimensions of the samples, $p$, as $O(n^2p+n^3)$~\cite{SVD2}. For a RHEED video with resolution of 1080x1080 pixels, frame rate of 5~frames per second (fps), and a duration of 25~minutes, the time needed for PCA to be completed on a single node of the Auburn University Easley cluster would be over a week. Thus, it is important to reduce both the number of frames sampled from a video and the dimensions of the video used in order to obtain a feasible run time. For the dimensionality, this is achieved by cropping the RHEED patterns down to a smaller subset of the recorded frame. During this process, a balance must be struck such that a sufficient amount of the pattern is retained and as much background as possible is excluded. The number of sample frames is reduced uniformly throughout the video by reducing the frame rate. The specific frame rate used is varied from growth to growth and is selected based on the growth rate of the material. If too low a frame rate is selected, information such as RHEED spot oscillations can be lost. In practice, it was found that for faster growths, a frame rate of 1~fps was sufficient while for slower growths a frame rate of 0.2~fps was sufficient. Additionally, blank frames at the beginning and end of growth were removed, as well as any frames that were partially obstructed during recording. With these modifications, the execution time for PCA on the sample described above was reduced to less than 2~hours. 

        In addition to the changes made to improve execution time, there are several image preprocessing algorithms that can improve image quality. The quality and sharpness of RHEED images are determined by not only the quality and structure of the film surface, but also the chamber geometry, chamber pressure, the quality of the phosphorous screen, and the quality of the RHEED camera. These additional effects do not reflect the quality of the film and can result in RHEED patterns that are not as clear as other samples, resulting in fainter features being obscured. Since the primary use of RHEED patterns is to be visually interpreted by a researcher, it is critical that these effects be mitigated as much as possible. In order to achieve this mitigation, three different intensity transformation functions are investigated.

        \begin{figure}
                \centering
                \includegraphics[scale=1]{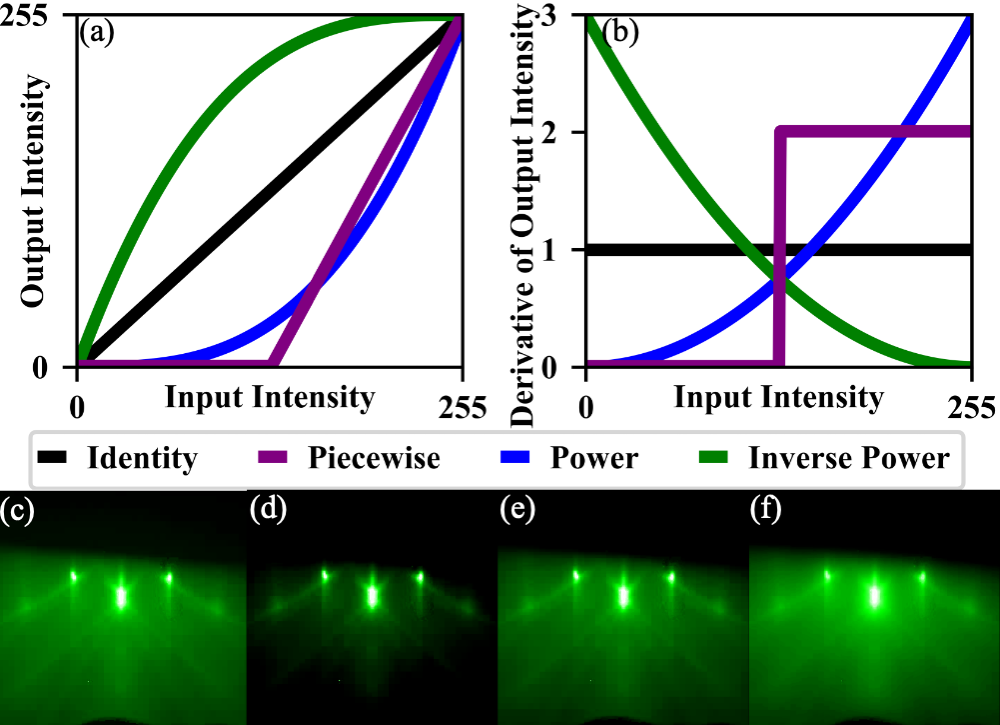}
            \caption{(a) The intensity transformation functions used on RHEED images. (b) The first derivatives of the intensity transformation functions. (c-f) An example LFO RHEED image and the effects of the (c) identity function, (d) piecewise, (e) power, and (f) inverse power.} 
            \label{fig:Background Changes}
        \end{figure} 
        
        Figure~\ref{fig:Background Changes} shows the three intensity transform functions and the identity function for reference. The piecewise intensity function is given by:

        \begin{equation}
            I_{out} = 
            \begin{cases} 
                0 & \text{if } I_{in} \leq T \\
                \frac{I_{in}-T}{1-T} & \text{if } I_{in} > T
            \end{cases}
        \end{equation}
        
        \noindent where $I_{in}$ and $I_{out}$ are the input and output intensities, respectively, and $T$ is the cutoff threshold. This piecewise function maps all pixel values below some specified background value to 0 and rescales the remaining pixel values between 0 and 255. In practice, the cutoff threshold is determined by calculating the distribution of pixel intensities for a given RHEED image and then determining a threshold pixel intensity from that distribution. This transformation results in a large area of the image becoming black while the remaining parts of the image have linearly increased contrast. 
        
        The power intensity function is given by:

        \begin{equation}
            \begin{aligned}
                I_{out} = I_{in}^{\gamma}
            \end{aligned}
        \end{equation}

        \noindent where $\gamma$ is a positive real number. This method uses a power function to rescale the pixel intensities rather than a linear one, resulting in reduced contrast between low intensity values and greater contrast between high intensity values. This is ideal for when the RHEED screen is heavily saturated, making it difficult to distinguish bright features from the background. The inverse power function results in the opposite effect, where low intensity values are increased in contrast while high intensity spots experience decreased contrast. The function is given by:

        \begin{equation}
            \begin{aligned}
                I_{out} = \lvert 1 - (1- I_{in})^{\gamma} \rvert.
            \end{aligned}
        \end{equation}

        \noindent This transformation is ideal for when there are weak features like Kikuchi bands and higher order spots that need to be enhanced to become visible. Figure~\ref{fig:Background Changes} shows an example RHEED image from an LFO sample and the results of the three transformation functions applied to it. 

        Due to the geometry of MBE chambers and slight differences in sample holder alignment on the stage, it is generally impossible to align samples the same way each time relative to the electron beam. Additionally, due to vibrations, magnetic interference, and mechanical noise present during growth, the alignment can shift during the course of a sample growth, either suddenly due to an instantaneous event such as the opening of a shutter or gradually due to sample drift on the stage. This causes a translation of the image during a RHEED recording and poses a problem for PCA and $k$-means as the changing locations of the pattern on the screen will be interpreted as a feature. In order to mitigate this, a system for identifying the specular spot of RHEED images and aligning them to the center of frame was devised. For a given RHEED image, the image is mirrored across a vertical axis and the difference between the mirror image and the original is calculated using a residual sum of squares. The mirror image is then translated 1 pixel horizontally and the residual sum of squares is calculated again. This is repeated for each possible horizontal alignment of the image and its mirror. RHEED patterns exhibit bilateral symmetry across the vertical axis and thus the difference between an image and its mirror will be minimized when aligned along the vertical axis of the specular RHEED spot. This method is highly effective at finding the horizontal center of a RHEED pattern and allows for different patterns to be aligned horizontally. This method will be referred to as the residual sum of squares (RSS) alignment.
        
        While RHEED images possess vertical mirror symmetry, RHEED images are not symmetric about the horizontal axis. However, the individual spots or streaks are symmetric about a horizontal axis. Thus in order to align the images vertically, the image is cropped and only the specular RHEED spot is used. First, an approximate center is found for the spot using a geometric mean of the pixel intensity. Using this approximated center, the image is cropped to the specular spot. Then, the RSS alignment algorithm is used to find the vertical center of the specular spot by repeating the process above using a horizontal mirror axis. In order to prevent background signal from dominating the RSS calculations, a low intensity filter is used that sets all values less than 90\% of the maximum spot intensity to 0. Thus, only the high intensity spots are used in the RSS alignment algorithm to determine the symmetry. Without this filter, the accuracy of this method drops considerably. 

    \subsection{PCA \& $k$-means}

        The PCA and $k$-means methods in this work follow the methods outline in \cite{Sydney}. The data preprocessing, PCA, $k$-means clustering, and analysis were implemented in Python 3.9.16. PCA and $k$-means were implemented using the scikit-learn package. The number of PCA components is specified by the user, and for this work it is set at 6, which retains $>99.98\%$ of the variance in the raw data. The resulting eigenvalues and eigenvectors are stored in h5 files using the h5Ppy package. The $k$-means clustering is performed on the results of the PCA with a number of clusters ranging from 1 to $k$ where $k$ is specified by the user and in this work is 10. Once $k$ gets beyond 10, the differences between clusters decreases and several clusters containing only a few frames begin to appear, making it hard to interpret the results. For each value of $k$, the clustering is run 100 times with different random starting configurations, in order to ensure an absolute ground state is found. Each clustering attempt is limited to 20 iterations, with the results being saved in h5 files. Graphs of the eigenvectors, eigenvalues, clustering distributions, and centroid images are created during post processing, which was also implemented in python.

\section {Results}

    \subsection{Intensity Rescaling}

        \begin{figure}[h!]
                \centering
                \includegraphics[width=3.375in]{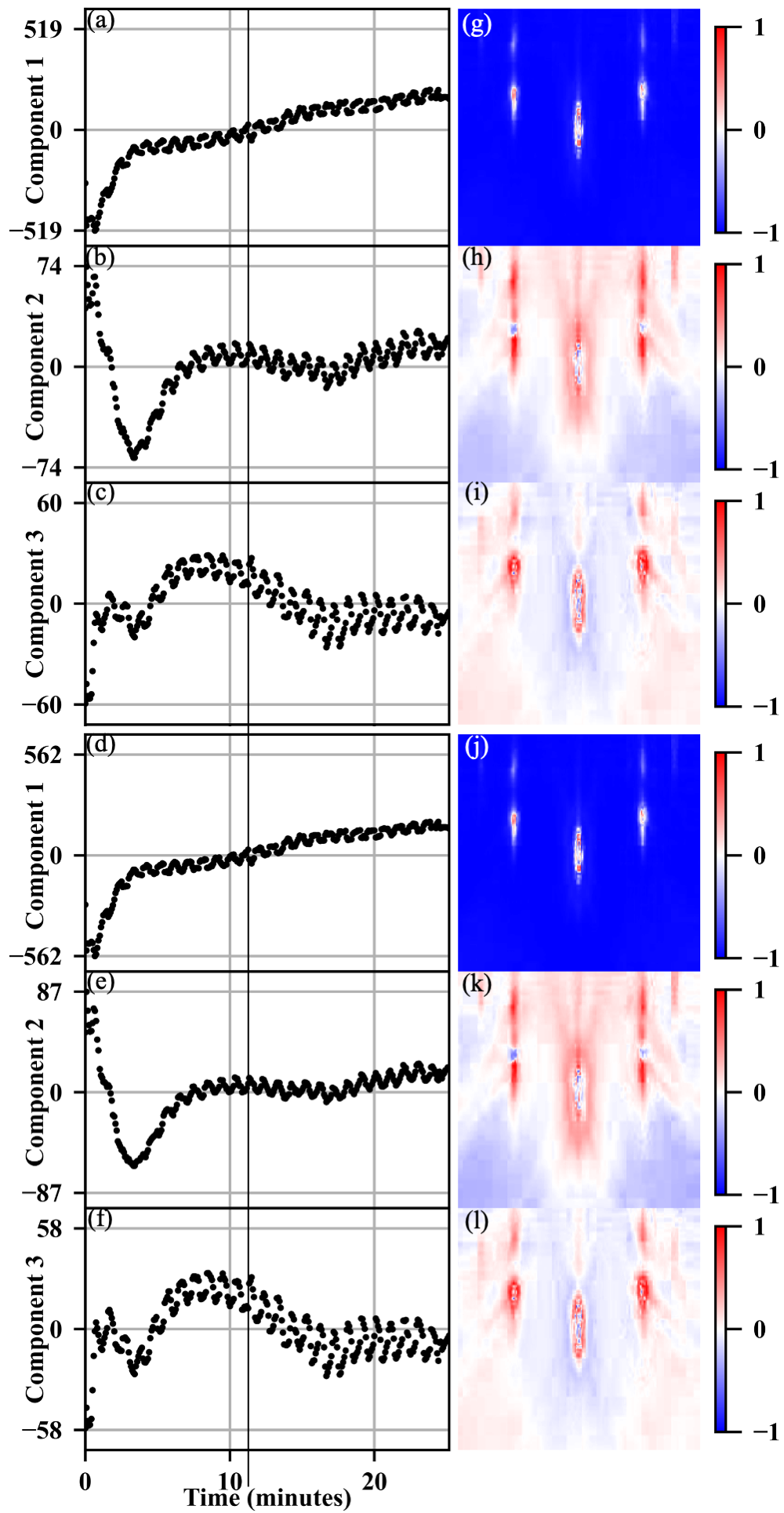}
            \caption{Results from the PCA of a recording containing the original RHEED recording of an LFO sample and the same recording after undergoing a power intensity transformation. (a-f) The eigenvalues plotted over the course of the growth and (g-l) the corresponding eigenvectors.} 
            \label{fig:PCA Power Transform}
        \end{figure} 
    
        \begin{figure}
                \centering
                \includegraphics[width=9cm]{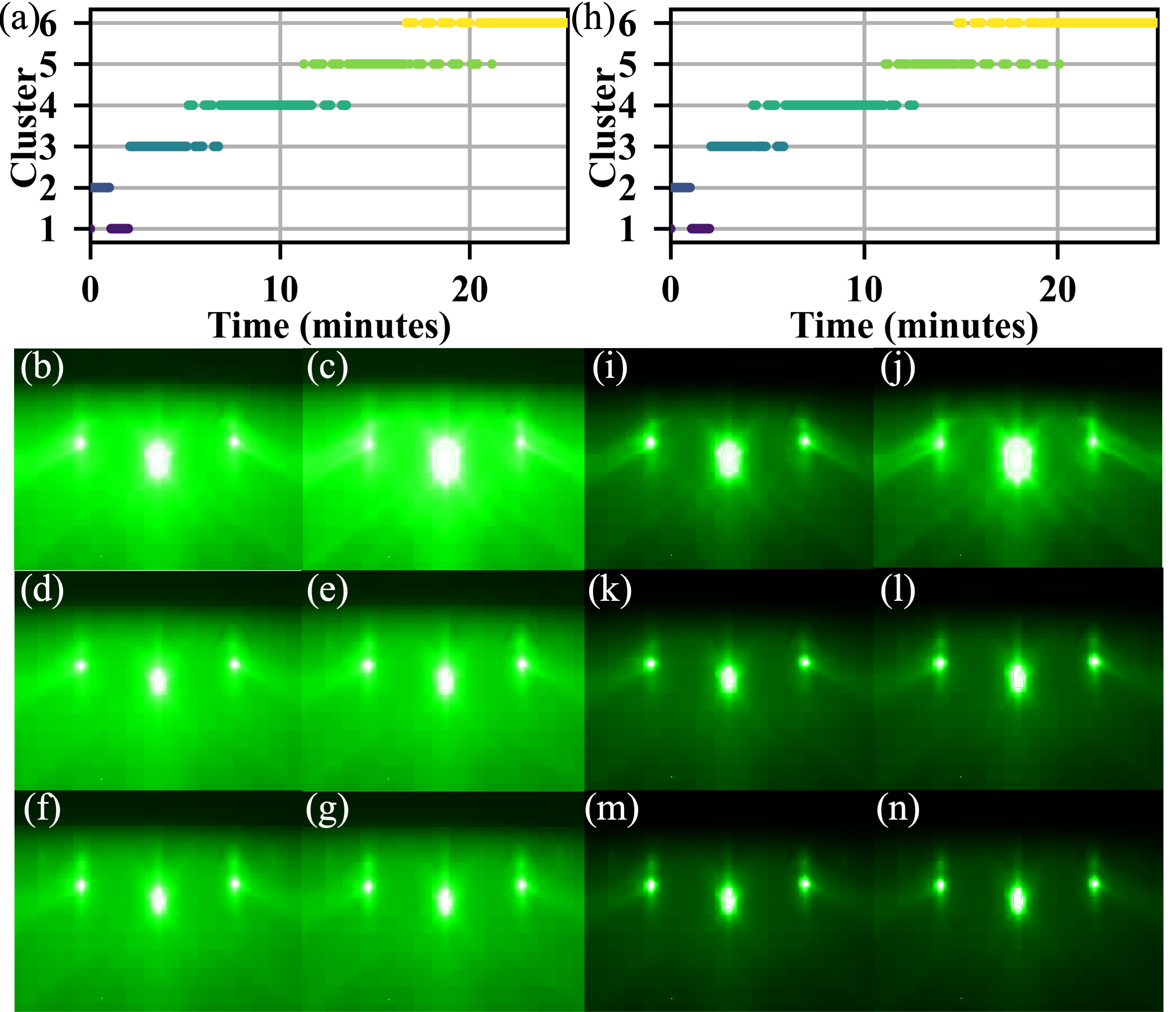}
            \caption{Results from the PCA of a recording containing the original RHEED recording of an LFO sample and the same recording after undergoing a power intensity transformation. Graphs of the clusters over the course of the growth for the (a) original and (h) transformed recordings. (b)-(g) The centroid images for the clusters plotted in (a). (i)-(n) The centroid images for the clusters plotted in (h).} 
            \label{fig:K-means Power Transform}
        \end{figure}

        Figures~\ref{fig:PCA Power Transform} and~\ref{fig:K-means Power Transform} show the results of PCA and $k$-means, respectively, for an LFO sample. Panels (a-c) and (g-h) of Figure~\ref{fig:PCA Power Transform} and (a-g) of Figure~\ref{fig:K-means Power Transform} correspond to the raw video collected during growth, while Panels (d-f) and (j-l) in Figure~\ref{fig:PCA Power Transform} and (h-n) in Figure~\ref{fig:K-means Power Transform} correspond to the same video after it has undergone the power transformation described in the methods section with $\gamma=2.0$. Despite the significant visual change in the RHEED patterns, there is no discernible difference between the original and transformed eigenvalues and eigenvectors. Due to the power recalling function, the contrast is toned down for lower intensity values and increased for higher intensity. As can be seen in the centroid clusters in Figure~\ref{fig:K-means Power Transform}, this improves the distinction between the Kikuchi bands, spots, and background. From the clustering plots in Figure~\ref{fig:K-means Power Transform}, we see there are no significant changes in the clusters, although there are some minor changes in the oscillations at the beginnings and endings of a few clusters. In cases like this, the improved contrast between the background and RHEED pattern features allows for improved visual analysis without affecting the PCA and $k$-means analysis.

    \subsection{RSS Alignment}

        \begin{figure}[h!]
                \centering
                \includegraphics[width=9cm]{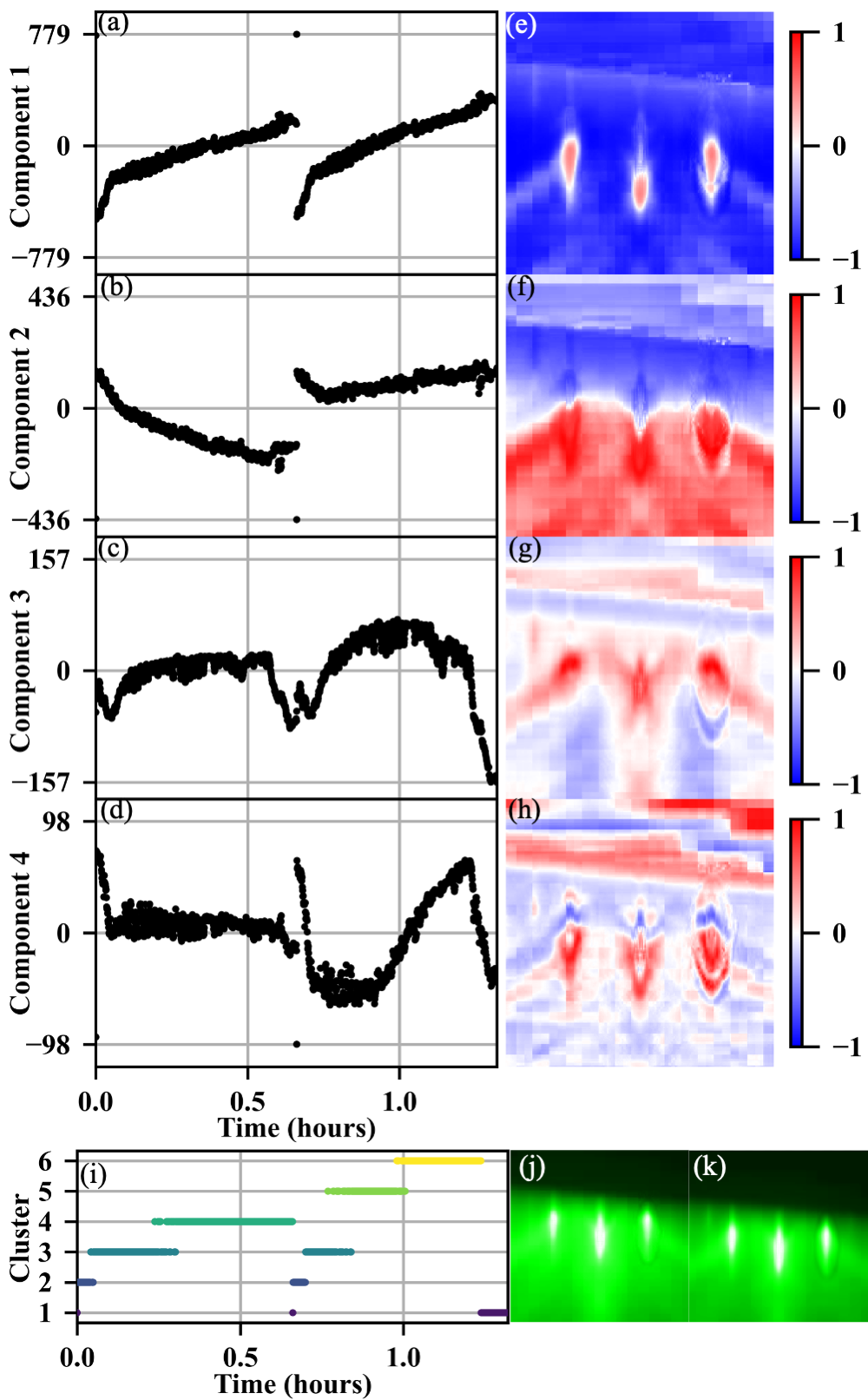}
            \caption{Results from PCA and $k$-means clustering of a recording containing the original RHEED recording of an LFO sample and the same recording with an artificial drift applied. (a-d)  The first four eigenvalues plotted over the course of the growths and (e-h) their corresponding eigenvectors. (i) A graph of the clusters over the course of the recordings and the centroid images of clusters (j) 3 and (k) 5, showing similar patterns that have undergone drift.} 
            \label{fig:K-means drift}
        \end{figure} 

        One of the common issues encountered during an extended growth is drifting or shifting of the RHEED pattern over time. This can be caused by mechanical vibrations present in the lab or can be the result of charging effects causing minor deflections in the electron beam. For the purpose of traditional RHEED interpretation, this issue is easily rectified by adjusting the alignment of the RHEED system; however, for numerical analysis such as specular intensity monitoring, PCA, and $k$-means, this is not the case. In PCA and $k$-means, small translations in the pattern result in different components and changes to the clustering that obscure the physical information present. This is seen clearly in Figure~\ref{fig:K-means drift}, which show the PCA and $k$-means results of an LFO sample grown using MBE. The first half of the graph corresponds to the original recording, while the second half of the graph corresponds to the same video that was translated 25 pixels right and 25 pixels down at a constant rate over the length of the video. From the clustering in Figure~\ref{fig:K-means drift}(i), it is clear that the drifting of the RHEED pattern causes the corresponding frames to be interpreted differently. The eigenvectors shown in Figure~\ref{fig:K-means drift}(e-j) shows that PCA is affected by this translation and constructs eigenvectors that can reproduce this translation. The eigenvalues in panels (c-d) show different behavior between the original and the drifting recording. The first appears to posses a vertical shift in intensity while retaining the overall progression; however, the second shows a drastic change in evolution while the third and fourth have taken on a periodic nature. 

        \begin{figure}[h!]
                \centering
                \includegraphics[width=9cm]{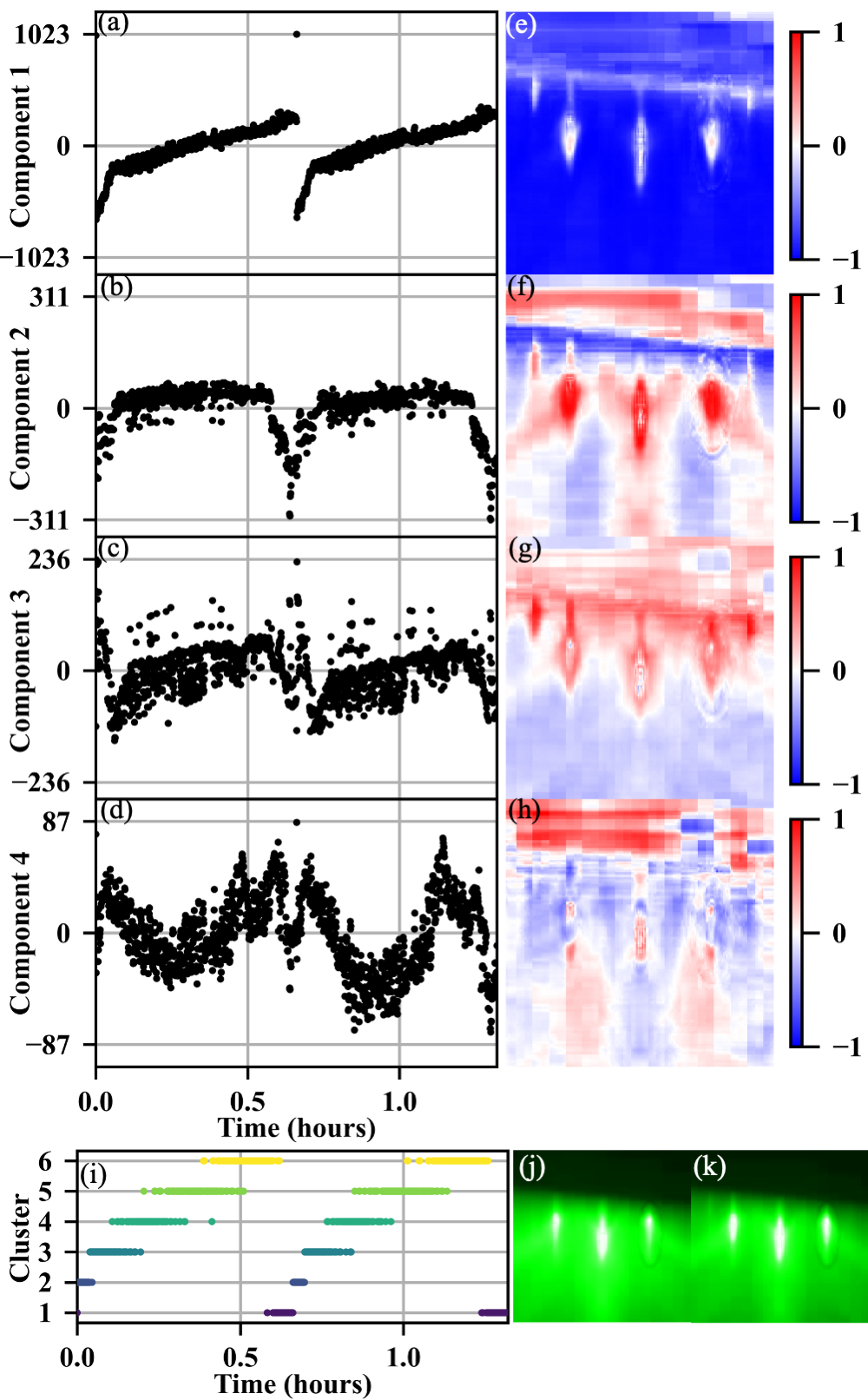}
            \caption{Results from PCA and $k$-means clustering of the same recording in Figure~\ref{fig:K-means drift} after the RSS alignment algorithm was applied. (a-d)  The first four eigenvalues plotted over the course of the growths and (e-h) their corresponding eigenvectors. (i) A graph of the clusters over the course of the recordings and the centroid images of clusters (j) 3 and (k) 5, showing differences in the centroid diffraction pattern after accounting for drift correction.} 
            \label{fig:K-means drift corrected}
        \end{figure} 

        In order to correct for drift or sudden shifts in the pattern, the RSS alignment algorithm described in the methods section is used. Figure~\ref{fig:K-means drift corrected} shows the PCA and $k$-means results for the same recording in Figure~\ref{fig:K-means drift} after the RSS alignment was applied. The clustering graph now shows nearly identical clustering for the original and drift recordings. The RHEED images in~\ref{fig:K-means drift corrected}(j) and (k) show that the centroid images of clusters 2 and 5 are now aligned with each other. The eigenvalue plots of the first 4 components show that the PCA now yields the nearly identical results for both recordings and as a result the eigenvectors now show features like the Kikuchi bands that correspond to the sample surface rather than focusing on the translation. From these results it is clear that the RSS alignment algorithm is able to successfully mitigate the effects of pattern translation and works to ensure PCA and $k$-means results are focused on significant features that inform on the film surface quality.

        \begin{figure}[t]
                \centering
                \includegraphics[width=9cm]{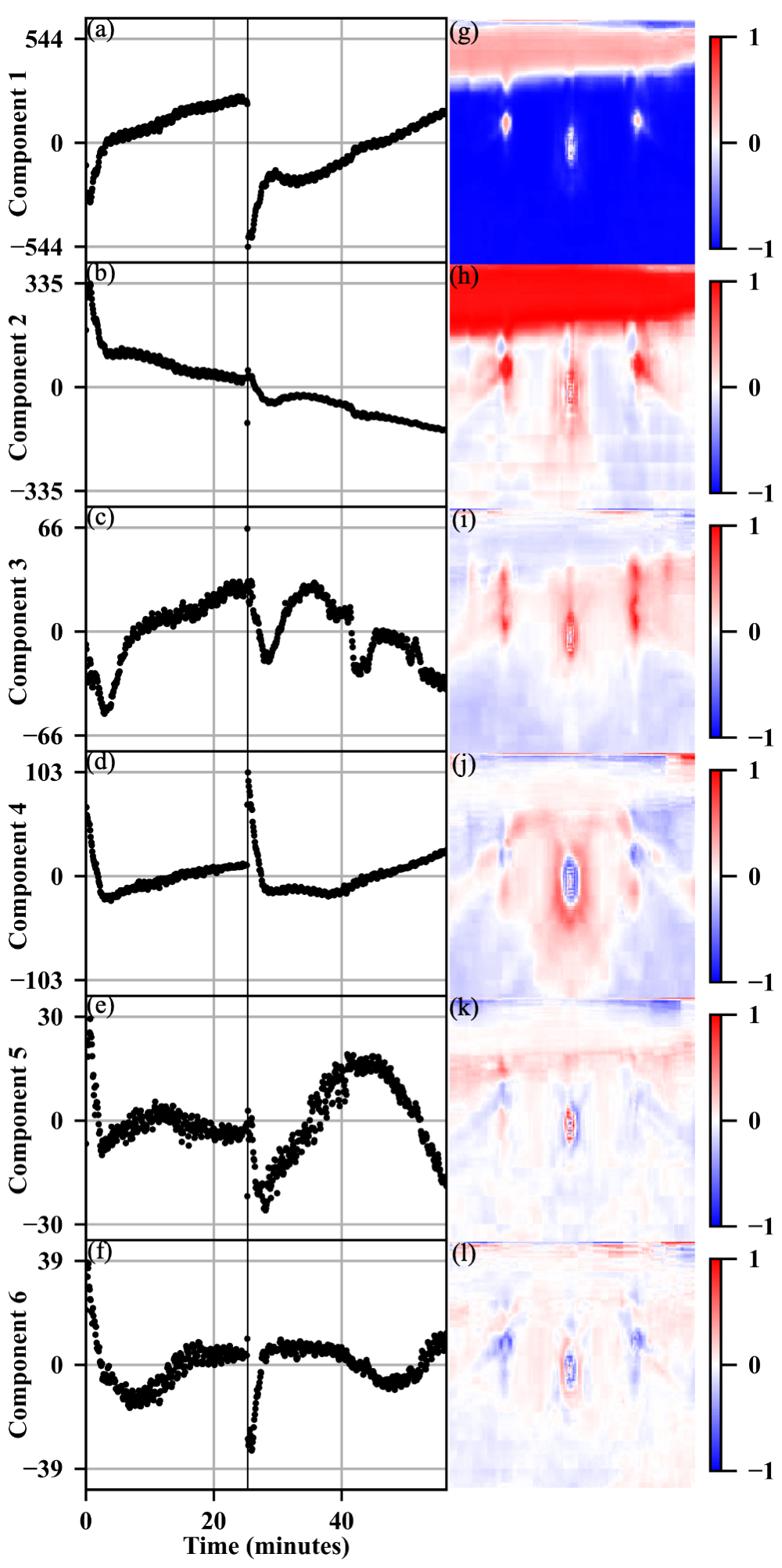}
            \caption{Results from the PCA of a recording containing 2 LFO samples. (a-f) The eigenvalues plotted over the course of the growths and (g-l) the corresponding eigenvectors.} 
            \label{fig:PCA Double LFO}
        \end{figure}

        \begin{figure}
                \centering
                \includegraphics[width=9cm]{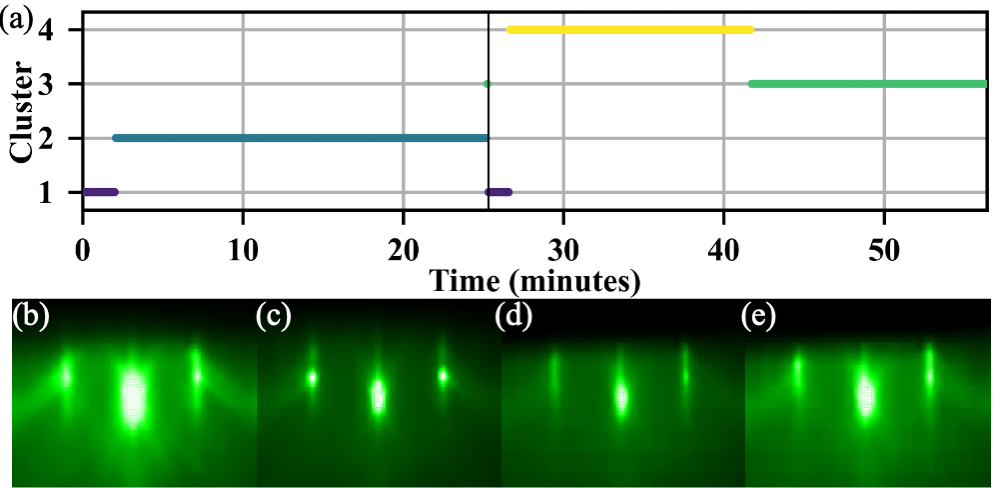}
            \caption{Results from the $k$-means clustering of a recording containing 2 LFO samples. (a) A graph of the clusters over the course of the growths; (b-e) The centroid images corresponding to the clusters in (a) in ascending numerical order left to right.}
            \label{fig:K-means Double LFO}
        \end{figure} 
        
        In addition to correcting drift and sudden shifts in alignment, RSS alignment is also able to align recordings from different samples with accuracy within a couple of pixels. This works best for videos that begin with clean substrates that possess high-quality RHEED patterns, as they will have the sharpest features. Figures~\ref{fig:PCA Double LFO} and~\ref{fig:K-means Double LFO} show the PCA and $k$-means results for the combined recordings of two LFO-Nb:STO samples. The samples were grown at identical conditions on the same day, with the first lasting 1500~s and the second lasting 1900~s. The two growths showed qualitatively different RHEED features in the final diffraction patterns, suggesting that drift in the La:Fe flux ratio had occurred over the course of the day. As expected, the eigenvalues are not the same for the two growths, though for a few components, such as 1, 3, and 4, the beginning of each growth is similar. The eigenvalues of the first component possess similar values and follow a similar progression. The corresponding eigenvector appears to represent the specular and first order spot intensity. The clustering graph in Figure~\ref{fig:K-means Double LFO} shows that the growths proceed through the same cluster initially for around 2~minutes, but diverge afterwards. The first cluster is highlighting the pattern of the Nb:STO substrate, which is replaced after $\approx120$~seconds. The first LFO sample then proceeds to cluster 2 which shows clear bands, although fainter than cluster 1, and distinct spots with minimal streaking. However, the second LFO sample procedes from cluster 1 to cluster 4 for $\approx900$ seconds before transitioning to cluster 3 where the RHEED is fainter and no Kikuchi bands are present. This softening along with less distinct spots suggest that islands have begun to form or parts of the film have begun to lose their structure and become amorphous. In this instance the two growths diverge into different results, however; without the RSS alignment, the direct comparison of the development the films surface would not be possible, as the PCA and $k$-means would have split the growths in the same manner observed in Figure~\ref{fig:K-means drift}. Using this approach can enable direct comparisons across numerous samples, highlighting relevant features that may vary or remain constant under slightly different growth conditions. Such differences could include growth temperature, varying stoichiometry such as the ratio of La to Fe in LFO due to drift in effusion cell fluxes, or differences in doping levels, for example, in La$_{1-x}$Sr$_x$FeO$_3$. Ultimately, the application of PCA and k-means to a large number of samples grown under differing growth conditions could allow new samples to be projected onto eigenvector images and k-means centroid images in real time during growth to objectively compare new samples to ideal and low-quality previous samples.

\section {Conclusion}

    Traditional RHEED analysis is primarily limited to specular spot intensity monitoring and qualitative analysis of the RHEED pattern. While this has served film growers well for decades, there is far more information contained in RHEED that can be teased out with the right tools. Here we have presented new tools to expand the power of PCA and $k$-means clustering in the analysis of RHEED videos. By analyzing the change in eigenvalues over the growth, new patterns, such as weak spot oscillations, are observed that were previously too weak to recognize. $k$-means clustering is able to utilize the eigenvalues to create mathematically significant clusters that distinguish between segments of the growth. RSS alignment of the images can address issues with drift on the sample stage and differences in alignment between samples, allowing for comparison of RHEED videos across samples in a quantitative fashion. This opens the door for use of unsupervised learning during growth, where real-time videos could be compared against past films to systematically evaluate film quality and adjust growth conditions on the fly.

\begin{acknowledgments}
P.T.G. and R.B.C. gratefully acknowledge funding for RHEED analytics from the National Science Foundation Division of Materials Research under award DMR-2045993. P.T.G. also acknowledges support from the Department of Energy's Office of Science Graduate Student Research Program (DE-SC0014664). R.P. gratefully acknowledges funding for LaFeO$_3$ film synthesis from the National Science Foundation Division of Materials Research under award DMR-1809847. The machine learning research was supported by the Center for Nanophase Materials Sciences (CNMS) which is a US Department of Energy, Office of Science User Facility at Oak Ridge National Laboratory.
\end{acknowledgments}

The data that support the findings of this study are openly available in Zenodo at http://doi.org/10.5281/zenodo.14649215.

\bibliography{My_References}
\end{document}